\documentclass[12pt]{article}

\catcode`\@=11
\def\marginnote#1{}
\newcount\hour
\newcount\minute
\newtoks\amorpm
\hour=\time\divide\hour by60
\minute=\time{\multiply\hour by60 \global\advance\minute by-\hour}
\edef\standardtime{{\ifnum\hour<12 \global\amorpm={am}%
        \else\global\amorpm={pm}\advance\hour by-12 \fi
        \ifnum\hour=0 \hour=12 \fi
        \number\hour:\ifnum\minute<10 0\fi\number\minute\the\amorpm}}
\edef\militarytime{\number\hour:\ifnum\minute<10 0\fi\number\minute}
\def\draftlabel#1{{\@bsphack\if@filesw {\let\thepage\relax
   \xdef\@gtempa{\write\@auxout{\string
      \newlabel{#1}{{\@currentlabel}{\thepage}}}}}\@gtempa
   \if@nobreak \ifvmode\nobreak\fi\fi\fi\@esphack}
        \gdef\@eqnlabel{#1}}
\def\@eqnlabel{}
\def\@vacuum{}
\def\draftmarginnote#1{\marginpar{\raggedright\scriptsize\tt#1}}
\def\draft{\oddsidemargin -.5truein
        \def\@oddfoot{\sl preliminary draft \hfil
        \rm\thepage\hfil\sl\today\quad\mil922 itarytime}
        \let\@evenfoot\@oddfoot \overfullrule 3pt
        \let\label=\draftlabel
        \let\marginnote=\draftmarginnote
   \def\@eqnnum{(\theequation)\rlap{\kern\marginparsep\tt\@eqnlabel}%
\global\let\@eqnlabel\@vacuum}  }

\def\preprint{\twocolumn\sloppy\flushbottom\parindent 1em
        \leftmargini 2em\leftmarginv .5em\leftmarginvi .5em
        \oddsidemargin -.5in    \evensidemargin -.5in
        \columnsep 15mm \footheight 0pt
        \textwidth 250mmin      \topmargin  -.4in
        \headheight 12pt \topskip .4in
        \textheight 175mm
        \footskip 0pt
        \def\@oddhead{\thepage\hfil\addtocounter{page}{1}\thepage}
        \let\@evenhead\@oddhead \def\@oddfoot{} \def\@evenfoot{} }

\def\titlepage{\@restonecolfalse\if@twocolumn\@restonecoltrue\onecolumn
     \else \newpage \fi \thispagestyle{empty}\c@page\z@ 
        \def\thefootnote{\fnsymbol{footnote}} }

\def\endtitlepage{\if@restonecol\twocolumn \else  \fi
        \def\thefootnote{\arabic{footnote}} \setcounter{footnote}{0}}

\catcode`@=12
\relax

\def\bea{\begin{array}}
\def\bem{\begin{displaymath}}
\def\beq{\begin{equation}}

\def\eea{\end{array}}
\def\eem{\end{displaymath}}
\def\eeq{\end{equation}}

\def\s2w{\sin^2 \theta_W}

\relax
\def\be{\begin{equation}}
\def\ee{\end{equation}}
\def\ba{\begin{eqnarray}}
\def\ea{\end{eqnarray}}

\def\w{\wedge}
\def\d{{\rm d}}

\def\k{\kappa}
\def\r{\rho}
\def\a{\alpha}

\def\b{\beta}

\def\G{\Gamma}
\def\dd{\delta}
\def\D{\Delta}
\def\e{\epsilon}

\def\m{\mu}

\def\l{\lambda}

\def\s{\sigma}

\def\cA{{\cal A}}
\def\cF{{\cal F}}

\def\Ct{{\widetilde C}}
\def\Gt{{\widetilde G}}
\def\At{{\widetilde A}}
\def\Rt{{\widetilde R}}
\def\Rh{{\hat R}}

\def\et{{\widetilde\epsilon}}
\def\ot{{\widetilde\omega}}

\def\Ft{{\widetilde F}}
\def\St{{\widetilde S}}

\def\t{\theta}
\def\rt{{\tilde r}}

\def\IR{\relax{\rm I\kern-.18em R}}

\def\inv{^{\raise.15ex\hbox{${\scriptscriptstyle -}$}\kern-.05em 1}}

\def\be{\begin{equation}}
\def\ee{\end{equation}}
\def\ba{\begin{eqnarray}}
\def\ea{\end{eqnarray}}

\def\tr{\,{\rm tr}\,}

\def\a{\alpha}
\def\b{\beta}

\def\G{\Gamma}
\def\d{{\rm d}}
\def\dh{\hat\d}
\def\e{\epsilon}

\def\m{\mu}

\def\r{\rho}
\def\l{\lambda}

\def\k{\kappa}
\def\s{\sigma}
\def\o{\omega}
\def\f{\phi}

\def\O{\Omega}

\def\ks{{k \kern-.5em /}}
\def\es{{\e \kern-.4em /}}
\def\ds{{\partial \kern-.5em /}}
\def\Ds{{D \kern-.6em /}}

\def\inv{^{\raise.15ex\hbox{${\scriptscriptstyle -}$}\kern-.05em 1}}

\def\wt{\widetilde}

\relax

\renewcommand{\theequation}{\thesection.\arabic{equation}}
\begin{document}
\topmargin-2.4cm
%
%
%
%
\begin{titlepage}
\begin{flushright}
LPTENS-02/56\\ hep-th/0303243 \\ March 2003
\end{flushright}
\vskip 3.5cm

\begin{center}{\Large\bf Anomalies in 
M-theory on singular $G_2$-manifolds }
\vskip 1.5cm
{\bf Adel Bilal$^{1}$ and Steffen Metzger$^{1,2}$}

\vskip.3cm
$^1$ CNRS - Laboratoire de Physique Th\'eorique, 
\'Ecole Normale Sup\'erieure\\
24 rue Lhomond, 75231 Paris Cedex 05, France

\vskip.3cm
$^2$ Sektion Physik, Ludwig-Maximilians-Universit\"at\\ 
Munich,
Germany\\

\vskip.3cm
{\small e-mail: {\tt adel.bilal@lpt.ens.fr, 
metzger@physique.ens.fr}}
\end{center}
\vskip .5cm

\begin{center}
{\bf Abstract}
\end{center}
\begin{quote}
When M-theory is compactified on $G_2$-holonomy manifolds
with conical singularities, charged chiral fermions are present
and the low-energy four-dimensional theory is potentially 
anomalous. We reconsider the issue of anomaly cancellation, 
first studied by Witten. We propose a mechanism that provides
local cancellation of all gauge and mixed gauge-gravitational
anomalies, i.e. separately for each conical singularity. It is 
similar in spirit to the one used to cancel the normal bundle
anomaly in the presence of five-branes. It involves smoothly 
cutting off all fields close to the conical singularities,
resulting in an anomalous variation of the 3-form $C$
and of the non-abelian gauge fields present if there 
are also $ADE$ singularities. 
\end{quote}


\end{titlepage}
\setcounter{footnote}{0}
\setcounter{page}{0}
\setlength{\baselineskip}{.7cm}
\newpage
%
%

\section{Introduction\label{Intro}}

M-theory compactified on a smooth 7-manifold $X$ of 
$G_2$-holonomy gives rise
to four-dimensional ${\cal N}=1$ supergravity 
coupled to $b_2(X)$
abelian vector multiplets and $b_3(X)$ neutral 
chiral multiplets
\cite{PT}. The theory contains no charged chiral 
fermions and there are no non-abelian
gauge symmetries. Both phenomena are generated when 
$X$ possesses
conical and $ADE$-singularities \cite{ADE,Cvetic,AW,AchW}. 
In this case the
low-energy four-dimensional theory is potentially 
anomalous, but it
was argued in ref.\,\cite{Witten} that all anomalies 
cancel against
various ``inflow'' terms.

The basic examples of conical singularities are taken from
the asymptotics of the well-known {\it non}-compact 
$G_2$-manifolds
\cite{BS}. Of course, Joyce's construction 
\cite{Joyce} gives compact
$G_2$-manifolds, but no explicit example with conical 
singularities is known.

On the other hand, there are generalisations of 
$G_2$-holonomy manifolds which naturally
are compact and have conical singularities. 
Mathematically they have so-called weak $G_2$-holonomy,
and in many cases we can write down the metric explicitly
\cite{BM}.
Physically, they correspond to turning on a background 
value for the supergravity four-form field strength $G=\d C$,
thus creating a non-vanishing energy-momentum tensor 
which increases the curvature of $X$ and makes it compact. 
When done appropriately one still has ${\cal N}=1$ 
supergravity, but now in $AdS_4$. Doing quantum
field theory, in particular loop calculations in $AdS$ 
spaces is highly non-trivial, but one can still study 
anomalies and their cancellation, since they are 
topological in nature. 

The aim of this note is to reconsider the anomaly cancellation
mechanism for singular $G_2$-manifolds
outlined in \cite{Witten}, but insisting on local 
cancellation, i.e. 
separately for each conical singularity. In
particular, one has to be careful about the correct
interpretation when using Stoke's theorem to rewrite 
bulk integrals as a sum of boundary terms. This
is a general feature of anomaly cancellation through inflow
from the bulk as soon as one has several ``boundary'' 
components.
We show how local anomaly cancellation can be properly 
achieved by appropriate modifications of certain 
low energy effective interactions like e.g. the Chern-Simons 
and the Green-Schwarz terms of eleven-dimensional 
supergravity, much in the same 
way as required for the cancellation of the normal bundle 
anomaly in the presence of five-branes \cite{FHMM}. The
basic feature of these modifications is to smoothly
cut off all the fields when a conical 
singularity is approached. 
If there are ADE singularities which generate 
non-abelian gauge fields, this cut-off procedure
induces corresponding modifications of the additional
interactions present in this case. To study the
mixed gauge-gravitational anomalies we also cut off the 
fluctuations of the geometry. Since we 
study quantum theory in a given background, only 
the fluctuations around this background are cut 
off, not the background itself.  In any case, the 
relevant interactions $S_i$ then naturally split 
into a ``bulk'' part $S_i^{(1)}$ and a sum of terms 
$S_i^{(2,\a)}$ localised at the various singularities 
$P_\a$. While the  $S_i^{(1)}$ are invariant, 
the variation of each $S_i^{(2,\a)}$ cancels the 
corresponding anomaly at $P_\a$ locally.
This method to achieve local cancellation is rather 
general and powerful. Exactly the same 
mechanism can also be applied to discuss local
anomaly cancellation 
on weak $G_2$-holonomy manifolds with conical singularities
as constructed in \cite{BM}.

This paper is organized as follows: in section 2, we review the
geometrical setup and the anomalies due to the chiral fermions
present at the singularities. We remind the reader 
how global anomaly cancellation was shown in \cite{Witten} and
explain why local cancellation still remained to be proven.
In section 3, we introduce our procedure of cutting off the 
fields close to the singularities and show how this leads to 
local  cancellation of the gauge anomaly in the
abelian case. We also give a preliminary discussion 
of the cancellation of the mixed gauge-gravitational 
anomaly. Section 4 deals with the non-abelian case where 
the cut-off procedure is more complicated due to the 
non-linearities. We show how the $SU(N)^3$ and 
mixed $U(1)_i\, G^2$ anomalies indeed are all cancelled locally.
Finally, we complete the discussion of the mixed 
gauge-gravitational anomaly. We conclude in section 5.
In an appendix we briefly describe the compact weak 
$G_2$-manifolds with two conical singularities 
constructed in \cite{BM}. They provide useful 
explicit examples to have in mind throughout
the main text.

\section{Global anomaly cancellation for abelian gauge fields}
\setcounter{equation}{0}

Anomalies that arise upon compactification of M-theory on
$G_2$-manifolds with conical singularities were first 
analysed by
Witten \cite{Witten}. The well-known non-compact metrics 
\cite{BS}
are asymptotically, for large $r$, a cone on a compact 
six-manifold
$Y$ with $Y=S^3\times S^3$, $Y={\bf CP}^3$ or $Y=SU(3)/U(1)^2$. 
The metrics on these manifolds all depend on some scale
which we call $r_0$, and the conical limit is $r_0\ll r$.
Of course, there is no singularity since, for small 
$r\sim r_0$,
these metrics are perfectly regular. 
Mathematically, it is only in the limit $r_0\to 0$ that a 
conical singularity develops. However, if $r_0$ is as small as
the eleven-dimensional Planck length (or less) then, 
from the long-wave length limit of supergravity, 
the manifold {\it looks as if} it
had a conical singularity. Said differently, the curvature 
is of
order ${1\over r_0^2}$ and supergravity ceases to be a valid
approximation. It was argued \cite{AW} that generic 
singularities of {\it compact} $G_2$-manifolds are also 
conical. 

In the vicinity of a conical singularity $P_\a$ we can 
always introduce a local coordinate $r_\a$ such that
the metric can be written as 
\be\label{e0} 
\d s_X^2\simeq \d r_\a^2 + r_\a^2 \d s_{Y_\a}^2 
\ee 
with $\d s_{Y_\a}^2$ the metric on the compact
six-manifold $Y_\a$. 
A necessary condition for $\d s_X^2$ to
have $G_2$-holonomy is  Ricci flatness. This in 
turn implies that $Y_\a$ is an Einstein space with 
${\cal R}^{Y_\a}_{ab}=5\dd_{ab}$. In fact, $Y_\a$
has weak $SU(3)$-holonomy. Furthermore, the 
Riemann tensors of $X$ and $Y_\a$ are related as 
$R^{Xab}_{\ \ \ \ cd} \simeq {1\over r_\a^2} 
\left( R^{Y_\a ab}_{\ \ \ \ \ cd} -\dd_c^a \dd_d^b 
+ \dd_d^a \dd_c^b \right)$, $a,b,\ldots = 1, \ldots 6$, 
and there are curvature invariants of $X$ that diverge 
as $r_\a\to 0$. It was argued in
\cite{Cvetic,AW,AchW} that at each such singularity 
$P_\a$ there is a set $T_\a$
of four-dimensional chiral supermultiplets 
$\Phi_\s , \ \s\in T_\a$. (This set may be empty as 
is the case for $Y_\a=S^3\times S^3$.)
They carry charges with respect to 
the abelian\footnote
{
We denote the (real) abelian gauge field one-forms 
$\cA_j=(\cA_j)_\m \d x^\m$ and their field strengths 
$\cF_j=\d \cA_j$. They are normalised such that the 
covariant derivative acting on a fermion of charge 
$q_j$ is $\partial_\m + i\, q_j (\cA_j)_\m$.
} 
gauge group $U(1)^k$ that
arises from the Kaluza-Klein reduction of the three-form 
$C$. Note, however, that they need not be charged with 
respect to {\it all} $U(1)$ gauge fields.

These charged chiral multiplets give rise to four-dimensional
gauge and mixed gauge-gravitational anomalies
``at a given singularity $P_\a$''.  
We follow the conventions and notations of \cite{BManom}:
the anomaly is the anomalous variation of the (Minkowskian) 
effective action $\delta\G^{\rm 1-loop}=\int \hat I_4^1$ 
where 
$\d \hat I_4^1=\delta \hat I_5$ and $\d\hat I_5=\hat I_6$. 
For a spin-${1\over 2}$ fermion of positive (Minkowskian) 
chirality we have 
\be\label{index}
\hat I_6 = - 2\pi \left[ \hat A(M_4) \,
{\rm ch}(F)\, {\rm ch}(i\, q_j \cF_j)\right]_6 \ ,
\ee 
where $F$ denotes a non-abelian field strength to be 
discussed below,
${\rm ch}(F) = \tr e^{{i\over 2\pi} F}$ and
$ {\rm ch}(i\, q_j \cF_j)=e^{-{1\over 2\pi} q_j\cF_j}$
denote the Chern characters of the non-abelian and 
abelian gauge groups,
and $\hat A=1+{1\over (4\pi)^2} {1\over 12} \tr R\w R + \ldots 
= 1 -{1\over 24}\, p_1 + \ldots$ is the Dirac genus. The 
overall minus sign in $\hat I_6$ takes 
into account the fact that 
a fermion of positive chirality in the Minkowskian 
translates into a negative chirality fermion in the 
Euclidean (see \cite{BManom,AGG}). Anomaly cancellation 
through classical inflow then  requires a non-invariance 
of the classical (Minkowskian) action $S$ 
such that $\dd S + \int \hat I_4^1 = 0$.

We see that in the presence of abelian gauge fields only,
one has a pure gauge anomaly characterised by the 6-form
\be\label{e1} 
\hat I_{6,\a}^{\rm gauge}={1\over 6 (2\pi)^2}
\sum_{\s\in T_\a}  \left( \sum_{j=1}^k q^j_\s \cF_j\right)^3 
\ee  
where
$\s$ labels the various chiral multiplets  $\Phi_\s$ 
present at $P_\a$
and $q_\s^j$ is the charge of  $\Phi_\s$ under the 
$j^{\rm th}$ $U(1)$.
The same chiral multiplets also give rise, at each 
singularity, 
to a mixed gauge-gravitational anomaly characterised by
\be\label{e2} 
\hat I_{6,\a}^{\rm mixed}=-{1\over 24}
\sum_{\s\in T_\a}  \left( \sum_{j=1}^k q^j_\s \cF_j\right) p_1'
\ee
where $p_1'= - {1\over 8 \pi^2} \tr R\w R$ is the first 
Pontryagin 
class of the four-dimensional space-time $M_4$. In ref. 
\cite{Witten}, it was argued that these anomalies are 
cancelled locally, i.e. separately for each singularity, 
by an appropriate non-invariance of the Chern-Simons and 
Green-Schwarz terms of eleven-dimensional 
supergravity:
\ba\label{e3}
S_{\rm CS} 
&=& -{1\over 12\kappa_{11}^2} \int C\w G\w G 
= -{1\over 6}{T_2^3\over (2\pi)^2} \int C\w G\w G \ ,
\\
S_{\rm GS} &=& - {T_2\over 2\pi} \int C\w X_8 \ ,
\ea
where, for convenience, we replaced $\kappa_{11}$ by 
the membrane tension $T_2$, via the usual relation 
$T_2^3={(2\pi)^2\over 2\kappa_{11}^2}$.
Here $G=\d C$ and $X_8$ is the standard gravitational 
eight-form to be given below. With our conventions, the 
$C_{MNP}$-field has dimension 0, 
so that the 3-form $C$ and the 4-form $G$ both have 
dimension $-3$. In particular, $T_2\, C$ is dimensionless.

Explicitly, the Kaluza-Klein reduction of $C$ is
\be\label{e4}
T_2\, C=c + \zeta_n\w \a_n + \cA_i\w \o_i 
+\f_k\, \Omega_k + \ldots
\ee
where $\a_n$, $\o_i$ and $\Omega_k$ are harmonic 
1-, 2- and 3-forms on $X$, and $c$, $\zeta_n$, 
$\cA_i$ and $\f_k$ are massless 3-, 2-, 1-form and 
scalar fields on $M_4$. The dots stand for 
contributions of massive fields. In particular, 
one gets a four-dimensional abelian 
gauge field 
$\cA_i=\cA_{i\m} \d x^\m$ for every harmonic 
2-form $\o_i$ on  $X$. Indeed, the gauge symmetry
$\dd C =\d \Lambda$ with $T_2 \Lambda=\e_i\o_i + \ldots$
corresponds to a $U(1)^k$ gauge transformation 
$\dd \cA_i=\d \e_i$. Note that the 
standard dimension for a gauge field $\cA_\m$ is 1, 
so that the one-forms $\cA_i$ have dimension 0 and hence the 
$\o_i$ are dimensionless.
The Kaluza-Klein reduction of $C$ implies a similar 
reduction for $G=\d C$ which in particular contains a term
$\cF^i\w \o_i$.

Due to the conical singularities, one has to be a bit 
more precise about which class of harmonic 2-forms one is
interested in. Inspection of the kinetic term 
$\sim \int \d C\w {}^* \d C$ shows that one needs
square-integrable harmonic forms on $X$, i.e. forms
satisfying $\int_X \o_i \w {}^* \o_i < \infty$,
in order to get massless 4-dimensional fields with
finite kinetic terms. In particular, square-integrability
requires an appropriate $r_\a$ dependence as 
$r_\a\to 0$. 
As long as we are in a neighbourhood $N_\a$ of the conical
singularity $P_\a$ where (\ref{e0}) holds, we can adapt
the results of ref.\,\cite{BM} : every $L^2$-harmonic 
$p$-form $\f_p^{(Y_\a)}$ on $Y_\a$ with $p\le 3$ 
trivially extends to a
harmonic $p$-form $\f_p^{(X)}=\f_p^{(Y_\a)}$ on $X$
such that $\int_{N_\a} \f_p^{(X)} \w {}^* \f_p^{(X)}$
is convergent. The same 
obviously is true for the Hodge duals ${}^* \f_p^{(X)}$ 
that are harmonic $q$-forms on $X$ with $q\ge 4$. In 
order to decide whether these forms are $L^2$ on $X$, 
i.e. whether $\int_X \f_p^{(X)} \w {}^* \f_p^{(X)}$ 
converges at all singularities, one needs more 
information about the global structure of $X$, 
which we are lacking. However, for
the examples of {\it weak} $G_2$ 
cohomogeneity-one metrics with two conical 
singularities constructed in \cite{BM} this
global information is available and
it was shown that the harmonic 
forms considered above are indeed $L^2$ and they 
are the only ones, so that
$b^p(X)=b^{7-p}(X)=b^p(Y_\a)$ for $p\le 3$. Also, 
since for a compact Einstein space of positive 
curvature like $Y_\a$ the first Betti number always 
vanishes, in these examples one then has $b^1(X)=0$. 
In the present 
case, we will simply need to assume that the 
$L^2$-harmonic $p$-forms on $X$ for $p\le 3$ are given, 
in the vicinity of $P_\a$, by the trivial extensions 
onto $X$ of the $L^2$-harmonic $p$-forms on 
$Y_\a$. In particular, then $k=b^2(X)=b^2(Y_\a)$ 
and $b^1(X)=0$. (Of course, for compact {\it smooth} 
$G_2$-holonomy manifolds $b^1(X)$ always vanishes.)
Hence, the gauge group is $U(1)^k$ and the 
terms $\zeta_n\w\a_n$ are absent in (\ref{e4}).

In \cite{Witten} it is argued that supergravity ceases
to be valid close to the singularities and 
$S_{\rm CS}$ and $S_{\rm GS}$ should be taken as
integrals over $M_4\times X'$ only, where $X'$ is $X$ with 
small neighbourhoods of the singularities excised. Then 
the boundary of $X'$ is $\partial X' =-\cup_\a Y_\a$,
and the variation of $S_{\rm CS}$ can be rewritten as a 
sum of boundary terms \cite{Witten}:
\be\label{e5}
\dd S_{\rm CS} \sim
\int_{M_4\times X'} \d \Lambda\w G\w G 
=-
\sum_\a \int_{M_4\times Y_\a} \Lambda\w G\w G \ .
\ee
Upon doing the KK reduction this yields
\be\label{e6}
\dd S_{\rm CS}
\sim \sum_\a \int_{M_4}\e_i\, \cF_j\w \cF_k  
\ \times \ \int_{Y_\a}\o_i\w \o_j\w \o_k \ .
\ee
Next one uses the relation
\be\label{e7}
\sum_{\s\in T_\a} q_\s^i  q_\s^j q_\s^k 
= \int_{Y_\a}\o_i\w \o_j\w \o_k 
\equiv d_{ijk}^{(\a)}\ ,
\ee
found to be true
for the three standard $Y_\a$ considered.
For the example of $Y_\a={\bf CP}^3$
there is a single harmonic 2-form $\o$, given in terms 
of the K\"ahler form $K$ as $\o={K\over \pi}$ and 
normalised such that $\int_{{\bf CP}^3} \o\w\o\w\o=1$.
This matches with the existence of a single multiplet with
$q=1$. Note that the orientation of $Y_\a$ 
is important. In the examples discussed in the 
appendix one has e.g. $Y_1={\bf CP}^3$ and 
$Y_2=-{\bf CP}^3$, so that one must have 
$q_1=1$ and $q_2=-1$. Given eq. (\ref{e7}), 
it was concluded in \cite{Witten} that eq. (\ref{e6}) 
cancels the gauge anomaly of the chiral multiplets 
(\ref{e1}).

This cannot be the full story, however. In eq. (\ref{e5}) 
one uses Stoke's theorem to rewrite a bulk integral 
as a {\it sum} of boundary contributions. While 
mathematically perfectly correct, it is not 
necessarily meaningful to assign a physical 
interpretation to the boundary contributions 
{\it individually}.\footnote{
As a trivial example consider integrating 0 over an interval
$[a,b]$. If $c(x)$ is {\it any} constant function we have
$\int_a^b 0=\int_a^b  \d c=c(b)-c(a)$. Obviously,
there is no meaning in assigning a value $c$ to the 
upper boundary and $-c$ to the lower one.
} 
These remarks suggest that the above argument
(\ref{e5}) - (\ref{e7}) is insufficient to show
the {\it local} character of the 
anomaly cancellation, separately at each singularity. 
Indeed, as it stands, the KK reduction of the
integrand of the l.h.s. of eq. (\ref{e5}) does 
not give the desired 
contribution. For a $U(1)^k$ gauge transformation with 
$T_2 \Lambda=\e_i \o_i$ the only piece contained
in $\d\Lambda \w G\w G=\d\Lambda \w \d C\w\d C$ which
is a 4-form on $M_4$ and only involves the massless fields
is $\d\e_i \w \cF_j\w \d\f_k$. In particular, the desired
piece $\d \e_i\w \cF_j\w \cF_k$ is a 5-form on $M_4$ and 
cannot contribute. We conclude that eq. (\ref{e5}) is a 
somewhat artificial rewriting of zero, at least for
the terms of interest to us, and that equations
(\ref{e6}) and (\ref{e7}) only prove global 
anomaly cancellation, i.e. cancellation after summing 
the contributions of all singularities $P_\a$.
Indeed, global cancellation of the anomaly is 
the statement that $\sum_\a \hat I_{6,\a}^{\rm gauge}=0$. As 
remarked in \cite{Witten}, this is a simple 
consequence of eq. (\ref{e7}) and 
$\sum_\a \int_{Y_\a} \o_i\w\o_j\w\o_k
=-\int_{X'} \d (\o_i\w\o_j\w\o_k)=0$.
However, local cancellation still remains 
to be proven. As we will show next, it will require a 
modification of $S_{\rm CS}$, much as when 
five-branes are present \cite{FHMM}.

\section{Local anomaly cancellation for abelian gauge fields}
\setcounter{equation}{0}

\subsection{The modified fields}

In the treatment of ref. \cite{FHMM} 
of the five-brane anomaly a small neighbourhood 
of the five-brane is cut out creating a boundary (analogous 
to $M_4\times Y_\a$). Then the anomalous Bianchi identity 
$\d G \sim \dd^{(5)}(W_6)$ is smeared out around this 
boundary and the $C$-field gets an anomalous variation 
localised on this smeared out region. Alternatively, this 
could be viewed 
as due to a two-form field $B$ living close to the boundary
and transforming as $\delta B=\Lambda$.
The CS-term is given by 
$\St_{\rm CS}=-{1\over 6} {T_2^3\over (2\pi)^2}
\int \Ct\w \Gt\w\Gt$ with 
appropriately modified $\Ct$ and $\Gt$ (which coincide 
with $C$ and $G=\d C$ away from the five-brane and its 
neighbourhood) such that $\dd\St_{\rm CS}$ is 
non-vanishing and cancels the left-over normal bundle 
anomaly.

Now we show that a similar treatment works for conical 
singularities. We first concentrate on the neighbourhood
of a given conical 
singularity $P_\a$ with a metric locally given by 
$\d s_X^2\simeq\d r_\a^2+r_\a^2 \d s_{Y_\a}^2$. 
The local radial 
coordinate obviously is $r_\a\ge 0$, the singularity 
being at $r_\a=0$. As mentioned above, 
there are curvature invariants of $X$ that diverge 
as $r_\a\to 0$. In particular, supergravity cannot be 
valid down to $r_\a=0$. Rather than cutting off the 
{\it manifold} at some $r_\a=\rt>0$, we cut off the 
{\it fields} which can be done in a {\it smooth} way. 
However, we keep fixed the geometry, and in particular 
the metric and curvature on $X$. 
Said differently, we cut off all fields that represent 
the quantum fluctuations but keep the background 
fields (in particular the background geometry) as before.
Introduce a small but finite regulator $\eta$ and
the regularised step function 
$\t_\a(r_\a-\rt)$ such that
\ba\label{e9}
\t_\a(r_\a-\rt) &=& 0
\qquad {\rm if} \quad 0\le r_\a\le \rt-\eta \ ,
\nonumber
\\
\t_\a(r_\a-\rt) &=& 1
\qquad {\rm if} \quad r_\a\ge \rt+\eta \ ,
\ea
with $\t_\a$ a non-decreasing smooth function 
between $\rt-\eta$ and $\rt+\eta$. (Outside the 
neighbourhood where the local coordinate $r_\a$ is 
defined, $\t_\a$ obviously equals 1.) We define 
the corresponding regularised $\dd$-function one-form 
as\footnote{
We write $\D$ rather than $\dd$ since the latter symbol 
already denotes the gauge variation of a quantity.}
\be\label{e10}
\D_\a=\d\t_\a \ .
\ee
Of course, if $X$ has several 
conical singularities (see the appendix for examples), 
$\t$ must have the appropriate 
behaviour (\ref{e9}) at each singularity $P_\a$. It can 
be constructed as the product of the individual $\t_\a$'s 
and then $\D$ becomes the sum of the individual $\D_\a$'s:
\be\label{dsum}
\D=\sum_\a \D_\a \ .
\ee 
When evaluating integrals one has to be careful since 
e.g. $\t^2\ne \t$, although 
$\t^2$ would be just as good a definition of a regularised 
step function. We write $\t^2\simeq\t$ which means that,
in an integral, one 
can replace $\t^2$ by $\t$ when multiplied 
by a form that varies slowly between $\rt-\eta$ and 
$\rt+\eta$. However, one has e.g. 
$\t^2\D=\t^2 \d\t ={1\over 3} \d \t^3 
\simeq {1\over 3} \d \t ={1\over 3} \D$ where a crucial 
${1\over 3}$ has appeared.  
Then, for any ten-form $\phi_{(10)}$, not containing 
$\t$'s or $\Delta$'s, we have
\be\label{e10bis}
\int_{M_4\times X} \phi_{(10)}\,  \t^n \D 
= \sum_\a {1\over n+1} 
\int_{M_4\times Y_\a}  \phi_{(10)} \ .
\ee
It is always understood that the regulator is removed, 
$\eta\to 0$, after the integration.

Now we cut off the fields with this $\t$ so that all 
fields vanish if $r_\a< \rt-\eta$ for some $\a$. 
Starting from $C$ and $G=\d C$ we define
\be\label{e11}
\hat C=C\t \ ,\qquad \hat G=G\t \ .
\ee
Then the gauge-invariant kinetic term for the 
$C$-field is constructed with $\hat G$:
\be\label{e17a}
S_{\rm kin}=-{1\over 4\k_{11}^2} \int \hat G \w {}^* \hat G
=-{1\over 4\k_{11}^2} \int_{r_\a > \rt} \d C\w {}^* \d C
\ee
and the $\cA_i$ resulting from the KK reduction of $C$
still are massless gauge fields.
To construct a satisfactory version of the Chern-Simons term,
we first note that,
of course, $\hat G\ne \d \hat C$ and 
$\d \hat G = G \D \ne 0$. However, we want a modified 
$G$-field which vanishes for  $r_\a< \rt-\eta$, is closed 
everywhere and is gauge invariant. Closedness is 
achieved by subtracting from $\hat G$ a term $C\w\D$, 
but this no longer is gauge invariant under 
$\dd C=\d\Lambda$. 

In order to maintain gauge invariance we add another 
two-form field $B$, 
that effectively only lives
on the subspace  $\rt-\eta<r_\a<\rt+\eta$, with
\be\label{e12}
\dd B=\Lambda \ .
\ee
In the limit $\eta\to 0$, $B$ really is a ten-dimensional 
field, although we treat it as an ``auxiliary'' field that has no 
kinetic term.
Of course, a gauge-invariant kinetic term could be added
as $\sum_\a \int_{M_4\times Y_\a} (C-\d B)\w {}^* (C-\d B)$
but it is irrelevant for our present purpose.
In any case
\be\label{e13}
\Gt=G\t - (C-\d B)\w\D
\ee
satisfies all requirements:
\be\label{e14}
\d\Gt=0 \ , \qquad \dd\Gt=0\ ,
\qquad \Gt=0 \quad {\rm for}\ r_\a<\rt-\eta \ .
\ee
We have
\be\label{e15}
\Gt = \d \Ct 
\ee
with 
\be\label{e15a}
\Ct=C\t + B\w \D 
\ee
and
\be\label{e15b}
\dd\Ct=\d \Lambda\, \t + \Lambda\w\D = \d ( \Lambda\, \t ) \ .
\ee

\subsection{The $U(1)^3$ anomaly}

All this is similar in spirit to ref. \cite{FHMM}, 
and we propose that $S_{\rm CS}$ should be replaced by
\be\label{e16}
\St_{\rm CS} = -{1\over 6} {T_2^3\over (2\pi)^2} 
\int_{M_4\times X} 
\Ct\w\Gt\w\Gt \ .
\ee
We may view the differences $\Ct-C$ and $\Gt-G$ as 
gravitational corrections in an effective low-energy 
description of M-theory. Actually, further gravitational 
terms of higher order certainly are present, but they are 
irrelevant to the present discussion of anomaly cancellation.

Note that in order to discuss local anomaly cancellation, 
i.e. cancellation singularity by singularity, we are 
not allowed to integrate by parts, i.e. use Stoke's theorem. 
More precisely, 
we must avoid partial integration in the $r$-direction 
since, as remarked above, this could shift contributions 
between the different singularities. However, once an 
expression is reduced to an integral over a given 
$M_4\times Y_\a$, corresponding to a given singularity, 
one may freely integrate by parts on $M_4\times Y_\a$, 
as usual. In particular consider any smooth $p$- and
$(9-p)$-forms $\varphi$ 
and  $\zeta$ not containing $\t$. Then
\ba\label{h11a}
\int_{M_4\times X} \d\varphi\w \zeta\ \t^n\ \D_\a 
&=&  {1\over n+1}\int_{M_4\times Y_\a} 
\d\varphi \w \zeta
\nonumber
\\
&&\hskip-3.cm =\ (-)^{p+1} {1\over n+1}\int_{M_4\times Y_\a} 
\varphi \w \d\zeta\
=\ (-)^{p+1} \int_{M_4\times X} \varphi\w \d\zeta\ \t^n\ \D_\a \ .
\ea
We see that whenever an integral contains a $\D_\a$ we are
allowed to ``integrate by parts'', but the derivative $\d$
does not act on the $\t$'s.

Writing out $\Gt$ and $\Ct$ explicitly, the modified 
Chern-Simons term reads
\ba\label{e15c}
\St_{\rm CS}  &=& 
-{1\over 6} {T_2^3\over (2\pi)^2} \int_{M_4\times X} 
\left[ C\w G\w G\, \t^3 
+ \Big( B\w G\w G - 2\, \d B\w C\w G\Big) \t^2 \D 
\right]
\nonumber
\\
\nonumber
\\
&\equiv& \St_{\rm CS}^{(1)} + \sum_\a \St_{{\rm CS}}^{(2,\a)} \ ,
\ea
where we used $\D=\sum_\a \D_\a$, see eq. (\ref{dsum}).
In the limit $\eta\to 0$, the first term 
$\St_{\rm CS}^{(1)}$ reproduces the usual bulk term, but only 
for $r_\a\ge \rt$, while the terms
$\St_{{\rm CS}}^{(2,\a)}$, due to the presence of $\D_\a$, 
each are localised 
on the ten-manifolds $M_4\times Y_\a$ close to the 
singularities $P_\a$. Although they look similar, they do
{\it not} arise as boundary terms. We have
\ba\label{e16a}
\St_{{\rm CS}}^{(2,\a)} &=&
-{1\over 18} {T_2^3\over (2\pi)^2}
\int_{M_4\times Y_\a} 
\left( B\w G\w G - 2 \d B\w C\w G\right) 
\nonumber
\\
&=& -{1\over 6}{T_2^3\over (2\pi)^2}
\int_{M_4\times Y_\a} B\w G\w G \ .
\ea
This result illustrates again eq. (\ref{h11a}).
An ``anomalous'' variation of each $\St_{{\rm CS}}^{(2,\a)}$ 
then arises since $\dd B=\Lambda \ne 0$:
\be\label{e16b}
\dd \St_{{\rm CS}}^{(2,\a)} =  
-{1\over 6} {T_2^3\over (2\pi)^2}
\int_{M_4\times Y_\a} \Lambda\w G\w G \ .
\ee
Of course, there is also the ``usual'' variation of
$\St_{\rm CS}^{(1)}$:
\be\label{e16c}
\dd \St_{\rm CS}^{(1)} =  
-{1\over 6} {T_2^3\over (2\pi)^2}
\int_{M_4\times X} \d\Lambda\w G\w G\ \t^3 \ .
\ee
{\it Globally}, this equals 
$-\sum_\a\dd \St_{{\rm CS}}^{(2,\a)}$
as could easily be seen when integrating by parts. Indeed, 
globally $\St_{\rm CS}$ is invariant. This is alright, 
since we know from \cite{Witten} that {\it globally}, 
i.e. when summed over the singularities, there are no 
anomalies to be cancelled.

Next, we will see what happens upon Kaluza-Klein reduction. 
We will keep all massless fields, not only the gauge fields.
In agreement with the above discussion we assume that 
$b^1(X)=0$. As before, let $\o_i$ be a basis of 
$L^2$-harmonic 2-forms and $\Omega_k$ of $L^2$-harmonic 
3-forms on $X$. Then
\ba\label{kk}
T_2\, C&=&c + \cA_i\, \o_i +\f_k\, \Omega_k + \ldots
\nonumber
\\
T_2\, G&=& \d c + \cF_i\,\o_i+\d\f_k\,\Omega_k + \ldots
\nonumber
\\
T_2\, B&=& \zeta + f_i\, \o_i + \ldots \ ,
\ea
where $f_i$ and $\f_k$ are  massless scalar fields  similar 
to axions, while $c$ and $\zeta$ are  3-form  and  2-form 
fields on $M_4$ respectively. The dots indicate contributions 
of massive fields. Under a ``gauge'' transformation with 
\be\label{gauge}
T_2\, \Lambda=\lambda+ \e_i\, \o_i + \ldots
\ee
one has for the 4-dimensional fields
\be\label{transf}
\dd c=\d\lambda\ , \quad
\dd \cA_i = \d \e_i \ , \quad
\dd \f_k=0 \ , \quad
\dd\zeta=\lambda \ , \quad
\dd f_i = \e_i \ .
\ee
Then in the ``bulk''-term
$\St_{\rm CS}^{(1)}$, the only non-vanishing contribution
of the massless fields is 
\be\label{bulk}
\St_{\rm CS}^{(1)} =  
-{1\over 6\, (2\pi)^2}
\int_{M_4} 3\, \f_k\, \cF_i\w \cF_j \ 
\int_X \Omega_k\w\o_i\w\o_j\, \t^3 + \ldots \ .
\ee
Obviously, this is gauge-invariant.
For the  terms $\St_{{\rm CS}}^{(2,\a)}$, localised
near the singularities $P_\a$, we get
\ba\label{e16d}
\St_{{\rm CS}}^{(2,\a)}=-{1\over 6\, (2\pi)^2}
&\Big( &
\int_{M_4} f_i\, \cF_j\w \cF_k \ 
\int_{Y_\a} \o_i\w \o_j\w \o_k 
\nonumber
\\
&-&\int_{M_4} \zeta \w\d\f_k\w\d\f_l \ 
\int_{Y_\a} \Omega_k\w\Omega_l \Big)
+ \ldots \ .
\ea
Under a $U(1)^k$ gauge transformation with $\lambda=0$ 
but $\e_i\ne 0$, the second term in this expression is 
invariant, but the first one is not.

We finally conclude that under a $U(1)^k$-gauge 
transformation with $\e_i\ne 0$ 
(but $\lambda=0$) we have
\ba\label{e23}
\dd\St_{\rm CS}^{(1)}&=&\ \ 0\ \ ,
\nonumber
\\
\nonumber
\\
\dd\St_{{\rm CS}}^{(2,\a)}
&=&
-\ {1\over 6 (2\pi)^2}
\int_{M_4}  \e_i\, \cF_j\w \cF_k \ 
\int_{Y_\a} \o_i\w\o_j\w\o_k  \ .
\ea
Using the relation (\ref{e7}) it is then 
obvious that, separately at each singularity,
this precisely cancels the gauge anomaly 
obtained from (\ref{e1}) via the descent equations.
Hence,
anomaly cancellation indeed occurs locally.

Before we go on, a remark is in order. To cancel the 
four-dimensional gauge anomalies we modified the 
eleven-dimensional Chern-Simons term, including a
new interaction with the ten-dimensional non-dynamical 
$B$-field. This was natural and necessary to have an 
invariant $\Gt$-field. As a result, the gauge variation 
of the KK reduction of
$\St_{\rm CS}^{(2,\a)}$ no longer vanishes and  
was seen to  cancel the four-dimensional 
gauge anomalies. This might look as if we had found a 
four-dimensional counterterm 
$\sim \int_{M_4} f_i\,  \cF_j\w \cF_k$ to cancel the anomaly.
Now, a relevant anomaly cannot be cancelled by the 
variation of a local four-dimensional counterterm of 
the gauge fields. The point is, of course, that 
this  not only contains the 
gauge fields but also the axion-like fields $f_i$ that 
arose from the non-dynamical $B$-field and it is the 
non-invariance of the $f_i$ that leads to anomaly 
cancellation. This is quite different from adding a 
four-dimensional counterterm of the gauge fields only.

It is also interesting to note that under transformations
with $\Lambda=\lambda$ we get non-vanishing 
$\dd \St_{{\rm CS}}^{(2,\a)}$, with no corresponding 
``fermion anomaly'' to be cancelled locally. Of course,
globally these variations vanish, but locally they do not.
However, this is not harmful as it would be for gauge anomalies,
since anyway, these transformations only affect
fields that do not propagate on $M_4$.

\subsection{The mixed gauge-gravitational anomaly}

The mixed gauge-gravitational 
anomaly (\ref{e2}) should be cancelled similarly through
local anomaly ``inflow'' from an appropriately modified
Green-Schwarz term. This now involves the gauge fields 
and the gravitational fields. Which fields should be cut 
off at the singularities? Our general philosophy is to 
keep fixed the background fields and in particular the 
background geometry, but to cut off the fluctuations around 
this background. If we call $\o_0$ the spin-connection 
of the background geometry, and $\o=\o_0+\s$ the one 
of the full geometry including the fluctuations $\s$
around $\o_0$, then one should cut off only $\s$ so that
$\o\equiv \o_0 + \s \to \wt\o\equiv 
\o_0 + \wt\s$. In principle one should then work 
with the gravitational 8-form $\wt X_8$ computed with 
this $\wt\o$, and start with
\be\label{g1}
\St_{\rm GS}=-{T_2\over 2\pi} \int \Ct \w \wt X_8 \ .
\ee
Dealing correctly with the cut-off spin connection requires
some machinery which we will only introduce in the next 
section where we deal with non-abelian gauge fields. However,
there we will also see that the two terms $\int \Ct \tr F^2$ and 
$\int \Ct \tr \Ft^2$ (with $F$ a non-abelian field 
strength and $\Ft$ its cut-off version) lead to exactly 
the same anomaly inflow for 
the mixed $U(1)_i\, G^2$ anomaly. Hence we expect that, 
in the same way, $\int \Ct X_8$ and $\int \Ct \wt X_8$ 
may also lead to the same anomaly inflow for the mixed 
$U(1)_i$-gravitational anomaly. We will explicitly 
verify this in section 4.4. Here we consider\footnote{
Alternative forms of the Green-Schwarz term would be
$-{T_2\over 2\pi} \int \Gt \w \wt X_7$ or
$-{T_2\over 2\pi} \int \Gt \w  X_7$. Although globally 
equivalent to (\ref{g1}), respectively (\ref{g2}), 
a priori they could lead to different local variations. 
Nevertheless, we have checked that the final result 
always equals (\ref{g7}).
}
\be\label{g2}
\St_{\rm GS}'=-{T_2\over 2\pi} \int \Ct \w X_8 \ .
\ee
As usual, $X_8$ is given by $X_8={1\over 24 (2\pi)^3} 
\left( {1\over 8} \tr R^4-{1\over 32} (\tr R^2)^2\right)$.

Let us comment on the sign of the Green-Schwarz term.
For the original Green-Schwarz term, written either as 
$\int C\w X_8$ or as
$\int G\w X_7$ with $\d X_7=X_8$ one can find 
both signs in the literature. With the convention we use, 
where the 
coefficient in  $S_{\rm CS}$ in front of $\int CGG$ is 
$-{1\over 6}\, {T_2^3\over (2\pi)^2}$, the 
correct the Green-Schwarz term is 
$-{T_2\over 2\pi}\int C\w X_8$.
In particular, having the same signs for the Chern-Simons
and the Green-Schwarz terms
is a necessary condition for the cancellation of both the 
tangent and the normal bundle anomalies of the five-brane
as described in ref. \cite{DLM,W5,FHMM}.
Similarly, as shown in \cite{BDS}, the correct Green-Schwarz 
term of the heterotic string with an 
$\hat X_8 \sim \left( {1\over 8} 
\tr R^4+{1\over 32} (\tr R^2)^2\right)+\ldots$
can only be reproduced from the interplay of the M-theory 
Chern-Simons and Green-Schwarz terms having the same 
signs. These issues where carefully reviewed in 
\cite{BManom,Metzger}.

In $X_8$ the curvature $R$ is evaluated with $\o=\o_0+\s$. 
In principal, one should consider arbitrary fluctuations 
$\s$ that do not necessarily preserve the product 
structure of the manifold, $M_4\times X$, but, for 
simplicity, we will assume they do.\footnote{
As always, dangerous anomalies are 
associated with the massless modes. Hence, we 
only need to consider fluctuations of the metric that are  
massless. Massless fluctuations that do not preserve 
the product structure would arise e.g. if $X$ had 
non-trivial Killing vectors. However, we know that for
$G_2$-manifolds this is not the case.
} 
Then one can rewrite 
$X_8$ in terms of the first Pontryagin classes 
$p_1'=-{1\over 8\pi^2}\tr R\w R\vert_{M_4}$ and 
$p_1''=-{1\over 8\pi^2}\tr R\w R\vert_{X}$
of $M_4$ and $X$ respectively as 
$X_8=-{\pi\over 48}\, p_1'\w p_1''$. 
Note that this is second and higher order in the fluctuations
since the background geometry of $M_4$ is flat ${\bf R}^4$.
Hence
\be\label{g3}
\St_{\rm GS}'=+{T_2\over 96} \int \Ct \w p_1' \w p_1'' \ .
\ee
Inserting the Kaluza Klein decomposition of $\Ct$ we get again
a ``bulk'' part and a sum of contributions localised 
close to the singularities:
\ba\label{g4}
\St_{\rm GS}'&=& \St_{\rm GS}^{'(1)} 
+ \sum_\a\St_{{\rm GS}}^{'(2,\a)} \ ,
\nonumber
\\
\St_{\rm GS}^{'(1)} &=&{1\over 96} \int_{M_4} \f_k\, p_1'\ 
\int_X \O_k\w p_1'' \ ,
\nonumber
\\
\St_{{\rm GS}}^{'(2,\a)} &=&{1\over 24} \int_{M_4} f_i\, p_1'
\ \ {1\over 4} \int_{Y_\a} \o_i\w p_1'' \ .
\ea
In the last integral over $Y_\a$, $p_1''$ now is the 
first Pontryagin class of $Y_\a$. This follows easily 
from the properties of the characteristic classes for 
the geometry at hand.\footnote{
Explicitly, this can be seen as follows.
On $X$ one has $\tr R\w R
= \sum_{\a,\b=1}^6 R^{\a\b}\w R^{\b\a}
+ 2 R^{\a 7}\w R^{7\a}$. But for the cones 
$R^{\a 7}=0$. Furthermore, the curvature 2-forms on $X$ 
and $Y_\a$ are related by $R^{\a\b}_X=R^{\a\b}_{Y_\a} 
- e^\a\w e^\b$ with $e^\a$ the 6-beins on $Y_\a$. Since 
with the relevant geometries we have
$R^{\a\b}_{Y_\a} \w e^\a\w e^\b=0$ one sees that
$\tr R\w R \vert_X = \tr R\w R \vert_{Y_\a}$ and 
hence $p''_1\equiv p_1(X)=p_1(Y_\a)$. 
Of course, eq. (\ref{g4}) is 
unchanged by the fluctuations of the geometry on $X$ since
$p''_1$ has topologically invariant integrals, and so does
$\o_i\w p''_1$.
}
Now one uses another relation which relates the charges
of the chiral fermions to the geometric properties of $Y_\a$
namely
\be\label{g5}
{1\over 4} \int_{Y_\a} \o_i\w p_1''(Y_\a)
= \sum_{\s\in T_\a} q^i_\s \ .
\ee
The only of the three examples for which this relation 
is non-trivial is $Y_\a={\bf CP}^3$ where
$p_1''=4\,\o\w\o$ and one correctly gets 
$\int \o\w\o\w\o=q=1$. Then
\be\label{g6}\St_{{\rm GS}}^{'(2,\a)} 
=\sum_{\s\in T_\a}{q^i_\s\over 24} \int_{M_4} f_i\, p_1'\ .
\ee
Finally, we conclude that under a $U(1)_i$ gauge 
transformation $\St_{\rm GS}^{(1)}$ is invariant while
\be\label{g7}
\dd \St_{{\rm GS}}^{'(2,\a)}  = 
\sum_{\s\in T_\a}{q^i_\s\over 24} \int_{M_4} \e_i\, p_1'\ .
\ee
This is exactly what we need to provide local cancellation
of the mixed gauge-gravita\-tional anomaly due to the chiral 
fermions associated with (\ref{e2}).

\section{Anomaly cancellation for non-abelian gauge fields}
\setcounter{equation}{0}

If $X$ has ADE singularities,
non-abelian gauge fields are generated.
Since ADE singularities have codimension four, 
the set of singular 
points is a three-dimensional submanifold $Q$. Such 
geometries have been discussed extensively in the 
literature see e.g. \cite{ADE,AW,Witten}. The interesting 
situation is when $Q$ itself has a 
conical singularity. In the neighbourhood of 
such a singularity $P_\a$, we may still think of $X$ 
as a cone on $Y_\a$, but now $Y_\a$ is an ADE orbifold. 
Let $U_\a$ be the two-dimensional singularity 
(fixed-point) set of $Y_\a$. Then locally $Q$ is a 
cone on $U_\a$.

On the seven-dimensional space-time $M_4\times Q$ there live 
non-abelian ADE gauge fields \footnote
{
Following the standard conventions used when dealing 
with anomalies of non-abelian gauge fields \cite{AGG}, 
we let $A=A^\a\, \l^\a$ with antihermitian 
$\l^\a$ : $(\l^\a)^\dagger = - \l^\a$. Then the 
covariant derivative is $\d+A$, and e.g. 
$\tr A\d A\d A$ is purely imaginary. This is in 
contrast with the abelian gauge fields $\cA_j$ which were 
taken to be real.
}
$A$ with curvature $F=\d A+A^2$.
After KK reduction they give rise to 
four-dimensional ADE gauge fields and
field strengths which we call again $A$ and $F$.
In addition, on $M_4$, we may still have abelian gauge fields
$\cA_i$ with field strength $\cF_i=\d \cA_i$, which arise from 
the KK reduction of the $C$-field.
The four-dimensional chiral supermultiplets $\Phi_\s$ 
present at the singularities $P_\a$ now couple to the 
gauge fields $A$ of the non-abelian group $G$ and are 
charged with respect to the abelian $\cA_i$. Then there 
are potentially $U(1)^3$, $U(1)\ G^2$ and $G^3$ anomalies. 
The first are cancelled as described above by inflow from 
the $\Ct\w\Gt\w\Gt$ term. The $G^3$ anomaly is present only
for $G=SU(N)$. 

\subsection{Consistent versus covariant anomalies}

In the non-abelian case anomalies can manifest themselves
in two different ways, as consistent or covariant anomalies
\cite{BZ}. As is well-known from the early days of the
triangle anomaly in four dimensions, if the regularisation
of the one-loop diagram respects Bose symmetry in the 
external gauge fields one gets the consistent anomaly. 
Alternatively one may preserve gauge invariance 
(current conservation) for two of the three external 
gauge fields (including contributions of square and 
pentagon diagrams), with all non-invariance only in 
the third field. This leads to the covariant form of 
the anomaly. For one positive chirality fermion 
this (integrated) covariant anomaly is given by the
following (real) expression
\be\label{e30}
{\cal A}_{\rm covariant}^{SU(N)^3}={i\over 2 (2\pi)^2}
\int_{M_4} \tr \e\,  F^2 \ .
\ee
Similarly, the (integrated) consistent anomaly 
for one positive chirality fermion is
\be\label{e29}
{\cal A}_{\rm consistent}^{SU(N)^3}={i\over 6 (2\pi)^2}
\int_{M_4} \tr \e\ \d\left( A\d A+{1\over 2} A^3\right) \ .
\ee
As recalled in section 2, 
the consistent anomaly is the anomalous variation of the
one-loop effective action 
${\cal A}_{\rm consistent}\equiv\delta \G^{\rm 1-loop}
=\int \hat I_4^1$ 
and is related via the descent equations\footnote{
$\tr F^3=\d \o_5$, $\dd\o_5=\d\o_4^1$, with
$\o_5=\tr\left( A F^2-{1\over 2} A^3 F +{1\over 10} A^5\right)
=\tr\left( A\d A\d A +{3\over 2} A^3\d A 
+{3\over 5} A^5\right)$ and
$\o_4^1=\tr \e\ \d \left( A F -{1\over 2} A^3\right)
=\tr \e\ \d\left( A\d A+{1\over 2} A^3\right)$.
}
to the invariant six-form 
${i\over 6 (2\pi)^2} \tr F^3=-2\pi [{\rm ch}(F)]_6$. 
Obviously, since the 
covariant anomaly cannot be obtained this way, 
it is not possible to find a local 
counterterm $\Gamma'$ of the gauge fields such that
${\cal A}_{\rm covariant}={\cal A}_{\rm consistent}
+\delta\Gamma'$. However, on the level of the corresponding
currents $J^\m$ one can find \cite{BZ} a local $X^\m$ such 
that $J^\m_{\rm covariant}=J^\m_{\rm consistent}+X^\m$.

Clearly, if the consistent anomalies cancel when summed 
over the contributions of all chiral fermions, the same 
is true for the covariant anomalies, and vice versa. 
Here, however, we want to cancel a non-vanishing anomaly 
due to chiral fermions (originating from a given conical 
singularity) by an appropriate anomaly inflow from a 
higher-dimensional action, i.e. by some $\delta S$. 
Such a setup respects Bose symmetry between all gauge 
fields and it is clear that we must cancel the consistent 
anomaly, not the covariant one. Indeed, when invoking 
anomaly inflow, one wants to show that the resulting total
effective action is invariant. But any non-invariance 
of part of the effective action must satisfy the 
Wess-Zumino consistency conditions \cite{ZW} and hence 
be the consistent anomaly. There has been some
discussion in the literature about consistent versus
covariant inflow \cite{Harvey1,Nakulich,Harvey2}: 
in all cases there is a consistent anomaly due to 
fermions to be cancelled by an inflow.\footnote{
In these papers a first inflow computation \cite{Harvey1}
gave a covariant anomaly inflow in discrepancy with the 
consistent anomaly due to the fermions.
It was then argued \cite{Nakulich} that a careful 
computation of the inflow actually gives two pieces 
for the current, 
the old covariant one, and a new one converting the 
consistent fermion current into a covariant one. 
However, a more fruitful interpretation is to
observe that the new inflow contribution is exactly 
what was needed to convert the old covariant inflow
into a consistent inflow.}

Similarly, for the mixed $U(1)_i\, G^2$ anomaly
the consistent form derives from the invariant 
6-form $-{1\over 2 (2\pi)^2} q^i \cF_i \tr F^2
=-2\pi\, [{\rm ch}(i q^i \cF_i)]_2\, [{\rm ch}(F)]_4$ via 
the descent equations. It can manifest itself as 
$-{1\over 2 (2\pi)^2}  q^i \e_i \tr F^2$ or
$-{1\over 2 (2\pi)^2}  q^i \cF_i \tr \e\, \d A$ or any 
combination of these two with total weight one:
\be\label{e31}
{\cal A}_{\rm consistent}^{U(1)_i G^2}
=-{q^i\over 2 (2\pi)^2} \int_{M_4} 
\left( \b\ \e_i \tr F^2 
+(1-\b)\ \cF_i \tr \e\, \d A \right) \ .
\ee
The parameter $\b$ can be changed by the addition
of a local four-dimensional counter\-term 
$\sim\int_{M_4} \cA_i\, \o_3(A)$ (where $\d \o_3(A)=\tr F^2$).
Note that the mixed $U(1)_i\, G^2$ anomaly is 
present for any $G$, not only $SU(N)$.
As for the pure $SU(N)^3$ anomaly, the mixed anomaly 
can also be expressed in a covariant form, but we
will not need it here.\footnote
{
As for the $SU(N)^3$ anomaly, the 
covariant form arises if, in the triangle diagram, one 
maintains $U(1)$ or $G$ gauge invariance at two of the 
vertices and then checks the gauge variations at the 
third vertex. If one probes for $U(1)$ invariance, there 
are non-abelian gauge fields at the two other vertices, 
while when probing $G$-invariance there are one abelian 
and one non-abelian gauge field at the other vertices, 
yielding a relative combinatorial factor 2. Hence the 
covariant mixed anomaly is
\be\label{e31a}
\nonumber
{\cal A}_{\rm covariant}^{U(1)_i G^2}=-{q^i\over 2 (2\pi)^2}
\int_{M_4} \left( \e_i \tr F^2 
+ 2 \cF_i \tr \e\, F \right) \ .
\nonumber
\ee
This is similar to (\ref{e31}) with $\b={1\over 3}$, 
but the coefficient is again 3 times larger and, of 
course, we have the covariant $ \cF_i \tr \e\, F$ instead 
of $ \cF_i \tr \e\, \d A$.
}

\subsection{The $SU(N)^3$ anomaly}

In ref. \cite{Witten} it was argued that these anomalies 
can be cancelled by the non-invariance of certain 
interactions, namely $S_1\sim \int_{M_4\times Q} K\w \o_5(A)$ 
for the $SU(N)^3$ anomaly and 
$S_2\sim \int_{M_4\times Q} C\w \tr F^2$ for the mixed 
one. Here $K$ is the curvature of a certain line bundle 
with a connection induced by the metric on $X$. 
$K$ is a 2-form on $Q$ and 
must obey $\d K=0$ except at the singularities of $Q$
where one could get $\dd$-function contributions. 
Thus it makes sense to define
\be\label{e38}
n_\a=\int_{U_\a} {K\over 2\pi}  \ .
\ee
Actually, ${K\over 2\pi}$ is the first Chern class 
of the line bundle and hence the $n_\a$ are integers.
To show that the variation of 
$S_1$ and $S_2$ cancel the fermion anomalies,
ref.\,\cite{Witten} again integrates by parts on $Q$.
According to our discussion above this only proves 
that anomalies cancel globally. To achieve local
cancellation we should first find interactions localised
close to the singularities, such that their variations 
individually cancel the fermion anomalies at each 
singularity. This will again involve cutting off the 
gauge fields using $\t$, but things will be slightly 
more complicated due to the non-linear structure of 
the non-abelian fields.

To see how we should cut off the non-abelian 
gauge fields we recall the important points of 
the abelian case: 1) all (fluctuating) fields should vanish 
close enough to the conical singularities and
equal the usual ones ``sufficiently far away'' 
from these singularities, and 
2) the modified field strengths should have the 
same properties as the unmodified ones. 
The first requirement allows fields to be a combination 
of terms involving $\t^n$ or $\t^{n-1}\D$ with $n\ge 1$. 
In the abelian case only $n=1$ occurred, see eqs 
(\ref{e13}) and (\ref{e15a}). The field strength 
$\Gt$ obeyed $\d\Gt=0$ and $\delta\Gt=0$ just as 
$\d G=0$ and $\delta G=0$. Hence, it satisfied also 
the second requirement. This was guaranteed because 
the relation between $\Gt$ and $\Ct$ was the same as 
the one between $G$ and $C$, namely $\Gt=\d\Ct$, while 
the gauge transformation was $\delta\Ct=\d(\Lambda\t)$ 
with an explicit $\t$ to make sure the transformed 
field also satisfies the first requirement.

Now we want to apply both requirements to the 
non-abelian case. The gauge field $A$ and field 
strength $F$ (defined on the 7-manifold $M_4\times Q$) should  
be replaced by cut off fields $\At$ and $\Ft$. It is 
then clear that the second requirement will be 
satisfied if
\be\label{f1}
\delta\At=\d \et+[\At,\et]
\ee
and
\be\label{f2}
\Ft=\d\At + \At^2 \ .
\ee
From the first requirement then
\be\label{f3}
\et=\e\, \t\ .
\ee
(Here, $\e$ is a Lie algebra-valued smooth function 
on $M_4\times Q$.) In particular these equations guarantee that
\be\label{f4}
\delta\Ft=[\Ft,\et] \quad , \qquad \d\Ft+[\At,\Ft] = 0 \ ,
\ee
as usual. The difference with the abelian case is that 
the non-linear structure (\ref{f1}) together with
(\ref{f3}) imply that $\At$ 
cannot simply be of the form $a\, \t+f\D$, but instead is
\be\label{f5}
\At=\sum_{n=1}^\infty 
\left( a_n\t^n + f_n \t^{n-1}\D\right) \ .
\ee
The $a_n$ are smooth 1-form fields on $M_4\times Q$, 
while the $f_n$ are smooth scalar fields, also on  
$M_4\times Q$ but effectively only on 
$M_4\times \cup_\a U_\a$. The latter are analogous 
to the $B$-field of the abelian case.
Note that ``in the bulk'', i.e. for $r_\a>\rt+\eta$ 
where $\t=1$ and $\D=0$, we have
\be\label{f6}
\At\Big\vert_{\rm bulk} =\sum_{n=1}^\infty a_n\ .
\ee
The gauge transformation (\ref{f1}) implies
\ba\label{f7}
\delta a_1=\d\e \qquad\quad &,& \quad \delta f_1=\e 
\nonumber
\\
\delta a_n= [a_{n-1},\e] \ \, &,& 
\quad \delta f_n = [f_{n-1},\e] \quad , \quad n\ge 2 \ .
\ea
In particular, it follows that the
``bulk''-field $\At\Big\vert_{\rm bulk}$ 
transforms as an ordinary gauge field,
\be\label{f8}
\delta \At\Big\vert_{\rm bulk}
=\d \e + \left[\ \At\Big\vert_{\rm bulk}\ ,\, \e\ \right]
\ee
as it should.
To simplify the notations below, we introduce
\ba\label{f9}
a&\equiv& a(\t)= \sum_{n=1}^\infty a_n \t^n 
\nonumber
\\
f&\equiv& f(\t)= \sum_{n=1}^\infty f_n \t^{n-1}
\ea
so that
\be\label{f10}
\At=a+f\D \ .
\ee
Let furthermore $a'\equiv\sum_{n=1}^\infty n\, a_n\, \t^{n-1}$,
as well as $\dh a\equiv \sum_{n=1}^\infty (\d a_n) \t^n$
and \break
$\dh f\equiv \sum_{n=1}^\infty (\d f_n) \t^{n-1}$.
Of course, $\dh$ behaves as an exterior derivative and,
in particular, $\dh^2=0$.
Then $\d a=\dh a-a'\D$ and
\be\label{f11}
\d\At=\dh a + (\dh f-a')\D \ .
\ee

Finally, we are in a position to show that the consistent 
fermion anomaly is cancelled by the non-invariance of the 
following interaction
\footnote{
Actually, we should start with an $\St_1$ where also 
$K$ is replaced by $\wt K\equiv K_0+\d\wt\s$ with $K_0$ 
corresponding 
to the background geometry and $\d\wt\s$ taking into 
account fluctuations around this background. Here 
$\wt\s$ is the appropriately cut-off spin connection 
on the line bundle: $\wt\s=\s\t+\rho\D$. It is easy 
to include this $\d\wt\s$-term into the computation 
which follows and show that it does not contribute 
to $\delta\St_1$.
}
\be\label{f12}
\St_1=-{i\over 6(2\pi)^2} \int_{M_4\times Q} 
{K\over 2\pi} \w \o_5(\At)
\ee
where $\o_5(\At)$ is the standard Chern-Simons 5-form 
with $\At$ replacing $A$, namely
\be\label{f13}
\o_5(\At)=\tr \left( \At\d\At\d\At + {3\over 2} \At^3\d\At
+{3\over 5} \At^5 \right) \ .
\ee
Note that with antihermitian $A$ (and $\At$), $\o_5(\At)$ 
is imaginary, while $K$ is real and, hence, $\St_1$ is 
real as it should in Minkowski space. The interaction
$\St_1$ is a sum of a ``bulk'' term not containing 
$\D$ and a term
linear in $\D=\sum_\a\D_\a$, so that we can again write
\be\label{f13a}
\St_1=\St_1^{(1)}+\sum_\a \St_{1}^{(2,\a)} \ ,
\ee
where 
the $\St_{1}^{(2,\a)}$ reduce to integrals over
$M_4\times U_\a$. Although it is straightforward to 
explicitly  compute the  $\St_{1}^{(2,\a)}$,
the resulting expressions are not very illuminating. 
However, their gauge variations turn out to be simple, and
this is why we will first compute the variations
$\dd \St_{1}^{(2,\a)}$, and
then reduce them to integrals over
$M_4\times U_\a$.

Since $\At$ satisfies the standard relation (\ref{f1})
we know that
\be\label{f14}
\delta \o_5(\At)= \d \o_4^1(\et,\At)
\ee
with
\be\label{f15}
\o_4^1(\et,\At) = \tr \et\, \d \left( \At \d\At 
+ {1\over 2} \At^3 \right) \ ,
\ee
so that
\be\label{f16}
\delta \St_1=-{i\over 6(2\pi)^2} \int_{M_4\times Q} 
{K\over 2\pi} \w \d \o_4^1(\et,\At) \ .
\ee
The next step is to explicitly evaluate the integrand,
substituting $\et=\e\,\t$ and (\ref{f10}) and (\ref{f11}) 
for $\At$ and $\d\At$. We get
\ba\label{f17}
\dd\St_1^{(1)} &=& 
-{i\over 6(2\pi)^2} \int_{M_4\times Q} 
{K\over 2\pi} \w
\tr \d\e \left( \dh a\, \dh a
+ {1\over 2} \dh a^3 \right) \t
\nonumber
\\
\dd\St_{1}^{(2,\a)}&=&
-{i\over 6(2\pi)^2} \int_{M_4\times Q} 
{K\over 2\pi} \w
\tr\Bigg(\e\, \dh a\, \dh a 
- \d\e\, (\dh a\, a' + a'\dh a)\, \t
+ \d\e\, \dh ( \dh a f + f \dh a) \, \t
\nonumber
\\
& &\hskip4.5cm +{1\over 2} \e\, \dh a^3 
-{1\over 2} \d\e\, (a' a^2 + a a' a + a^2 a') \t
\nonumber
\\
& &\hskip4.5cm +{1\over 2} \d\e\,\dh (a^2 f + a f a + f a^2)\, \t
\Bigg)\ \ \D_\a \ .
\ea
(Of course, $\dh a^3$ is shorthand for
$\dh a\, a^2 - a \dh a\, a + a^2 \dh a$, etc.)

To go further, we perform 
the Kaluza-Klein reduction. Each 1-form field $a_n$ on 
$M_4\times Q$ becomes a 1-form field on $M_4$ which we 
also denote by $a_n$, and a scalar field $\chi_n^l$ on $M_4$  
for every harmonic 1-form $\b_l$ on 
$Q$, plus massive modes. The scalars $f_n$ simply 
become scalars on $M_4$ (again denoted $f_n$), 
plus massive modes. 
Since $K\w\D_\a$ already is a 3-form on $Q$,
the $\chi^l\b_l$ cannot contribute in 
$\dd\St_{1}^{(2,\a)}$, while in $\dd\St_1^{(1)}$
we must pick out the part linear 
in $\chi^l\b_l$. It is 
$\tr \d\e\, \dh \left( \chi^l\dh a + \dh a\chi^l
+{1\over 2} a^2\chi^l -{1\over 2} a\chi^l a +
{1\over 2} \chi^l a^2\right) \b_l$, where now 
$\dh\to\d_4$ is the exterior derivative on 
$M_4$. Clearly, after integration over $M_4$ 
this term vanishes:
\be\label{f17a}
\dd\St_1^{(1)}=0\ .
\ee
It remains to evaluate $\dd\St_{1}^{(2,\a)}$ with $a_n$ 
and $f_n$ now 1- and 0-forms on $M_4$. 
To see how to perform the integrals over $Q$, 
consider e.g. the terms that only involve two $a$-fields:
\ba\label{f18}
& &\int_{M_4\times Q} {K\over 2\pi} \w
\tr\left(\e\, \dh a\, \dh a 
- \d\e\, (\dh a\, a' + a'\dh a)\, \t\right)\ \D_\a
\nonumber
\\
&=&
\int_{M_4\times Q} {K\over 2\pi} \w
\tr \sum_{n,m=1}^\infty 
\left(\e\, \d a_n\, \d a_m 
- \d\e\, \d a_n\, m\, a_m - \d\e\, n\, a_n\d a_m)\right)
\, \t^{n+m} \D_\a
\nonumber
\\
&=&\int_{M_4\times U_\a} {K\over 2\pi} \w
\tr \sum_{n,m=1}^\infty {1\over n+m+1}
\left(\e\, \d a_n\, \d a_m 
- m\, \d\e\, \d a_n\, a_m - n\, \d\e\, a_n\d a_m)\right)
\ .
\nonumber
\\
\ea
At this point the $Q$-integral is reduced to a sum 
of integrals over $M_4\times U_\a$ and now we can 
safely integrate by parts.
The three terms then all 
are $\e\, \d a_n\, \d a_m$ and the coefficients add 
up as ${1\over n+m+1} (1+m+n)=1$. Using (\ref{e38}) we get 
$n_\a \int_{M_4} \tr \e\, \d A\, \d A$, where now 
\be\label{f18a}
A=\sum_{n=1}^\infty a_n 
\ee
is the Kaluza-Klein reduction of the ``bulk'' field 
$\At\Big\vert_{\rm bulk}$ encountered before in
(\ref{f6}).
Similarly one sees that the terms involving $f$ 
do not contribute,\footnote{
Of course, we could have ``integrated by parts'' according 
to the rule  (\ref{h11a}) directly in $\dd\St_1^{(2,\a)}$
in (\ref{f17}), showing immediately that the $f$-fields 
do not contribute.
} 
while the terms involving three 
$a$-fields add up to give 
$n_\a \int_{M_4} \tr \e\, {1\over 2} \d A^3$.
We conclude that
\be\label{f19}
\dd\St_{1}^{(2,\a)}=-\, n_\a\ {i \over 6(2\pi)^2} 
 \int_{M_4} 
\tr \e\, \d \left( A\, \d A + {1\over 2} A^3 \right)
\ .
\ee
Provided the $n_\a$ coincide with the number of 
charged chiral multiplets present at the singularity 
$P_\a$, as suggested in \cite{Witten}, the 
non-invariance of the interaction (\ref{f12}) 
cancels the $SU(N)^3$ anomaly locally, separately 
at each singularity.

Quite remarkably,
the final result is simple with all contributions 
of the different $a_n$ adding up to reproduce 
$\o_4^1(\sum_n a_n)\equiv \o_4^1(A)$. Alternatively, 
one might have first expanded $\St_1$. Then 
$\St_{1}^{(2,\a)}$ would have reduced to an
integral over  $M_4\times U_\a$ with the
integral of ${K\over 2\pi}$ over $U_\a$ just giving
$n_\a$. The result would have been
a four-dimensional action involving infinitely 
many fields $a_n$ and $f_n$. While the gauge 
transformations of each term individually are 
complicated, we know that they sum up to give
(\ref{f19}).

\subsection{The $U(1)_i\, G^2$ anomaly}

It remains to discuss the cancellation of the mixed 
$U(1)_i\, G^2$ anomaly (\ref{e31}). 
Clearly, the variation of an interaction 
$\sim \int_{M_4\times Q} \Ct\w \tr F^2$ can
cancel the consistent anomaly (\ref{e31}) with $\b=1$. 
However, following our general philosophy, we 
should really start with
\be\label{e44}
\St_2 = {T_2\over 2(2\pi)^2} \int_{M_4\times Q} 
\Ct\w \tr \Ft^2 \ .
\ee
Note that in the bulk this coincides with the standard
interaction ${T_2\over 2(2\pi)^2}\int C\w\tr F^2$.
Since $\tr \Ft^2 $ was designed to be gauge invariant 
under (\ref{f1}) only $\delta \Ct$ contributes to the 
gauge variation of $\St_2$. 
Again, we write $\St_2=\St_2^{(1)}+\sum_\a \St_{2}^{(2,\a)}$.
Inserting 
$\delta \Ct=\d\Lambda\, \t+\Lambda\, \D$ from 
eq. (\ref{e15b}) and $\Ft=\dh a+a^2+(\dh f+a f-f a-a')\,\D$ 
from (\ref{f10}) and (\ref{f11}) we get
\ba\label{f21}
\dd\St_2^{(1)} &=& {T_2\over 2(2\pi)^2} 
\int_{M_4\times Q} 
\d\Lambda \ \tr(\dh a\dh a+2 a^2\dh a)\, \t
\nonumber
\\
\dd\St_{2}^{(2,\a)} &=& {T_2\over 2(2\pi)^2} 
\int_{M_4\times Q} 
\Big[ \Lambda \, \tr (\dh a\dh a+2 a^2\dh a)
\nonumber
\\
&& \hskip2.cm
+ 2\,\d\Lambda\, \tr \left( \dh (f\dh a+f a^2)
-a'(\dh a+a^2)\right)\t \Big]\ \D_\a \ .
\ea
When we perform the Kaluza Klein reduction, 
$T_2 \Lambda\to \e_i\, \o_i$ and 
$a_n\to a_n + \chi^l_n\b_l$, in $\dd\St_2^{(1)}$
only terms linear in $\b_l$ can 
contribute, but they vanish after partial integration 
over $M_4$, just as for $\delta \St_1^{(1)}$. Also as 
before, in $\dd\St_{2}^{(2,\a)}$, $\chi^l_n\b_l$ 
cannot contribute, while the terms containing $f$ 
again vanish after partial integration over $M_4$. 
The remaining terms combine to yield
\be\label{f22}
\dd\St_{2}^{(2,\a)} 
= {1\over 2(2\pi)^2}  \int_{U_\a} \o_i\
\int_{M_4} \e_i\, \tr(\d A \d A + 2 A^2 \d A) \ .
\ee
Provided 
\be\label{f22a}
\int_{U_\a} \o_i=\sum_{\s\in T_\a} q^i_\s \ ,
\ee 
this exactly cancels the mixed 
$U(1)_i\, G^2$ anomaly locally. Note that the variation of
$\St_2' = {T_2\over 2(2\pi)^2} \int_{M_4\times Q} 
\Ct\w \tr F^2$ would have produced exactly the same result.

\subsection{The mixed gauge-gravitational anomaly once more}

Now we dispose of the necessary machinery to show
that  the variation of the modified Green-Schwarz term 
(\ref{g1}) with cut-off $\wt X_8$ leads to the same local
anomaly contribution as the variation of (\ref{g2}) 
using the ordinary $X_8$.

To begin with, we replace the spin connection 
$\o=\o_0+\s$ by its cut-off version
\be\label{h1}
\ot=\o_0+\wt\s \ .
\ee
$\o_0$ represents the fixed background ${\bf R}^4\times X$
and $\s$ the fluctuations. For the time being we make 
no assumption about $\s$, but later on we will again 
restrict to fluctuations that preserve the product structure 
of the manifold. Again we write
\be\label{h2}
\wt\s=\eta(\t)+\r(\t) \D \equiv \eta+\r \D
\ee
with $\eta(\t)=\sum_{n=1}^\infty \eta_n\, \t^n$ and
$\r(\t)=\sum_{n=1}^\infty \r_n\, \t^{n-1}$. 
Of course,
\be\label{h2a}
\sum_{n=1}^\infty \eta_n = \s \ ,
\ee
since in the bulk, where $\t=1$ and $\D=0$, we want 
$\wt\s$ to coincide with $\s$.

We require that 
under a local Lorentz transformation with parameter 
$\e_{\rm L}$ one has
\be\label{h3}
\dd\ot=\d \et_{\rm L} + [\ot\, ,\, \et_{\rm L}] \quad , 
\quad \et_{\rm L} = \e_{\rm L}\t \ .
\ee
This ensures that 
\be\label{h4}
\Rt=\d\ot+\ot^2
\ee
transforms covariantly: $\dd\Rt=[\Rt,\et_{\rm L}]$ and
\be\label{h5}
\dd \tr \Rt^n=0 \ .
\ee
Comparing powers of $\t$ and $\D$ in eq. (\ref{h3}) shows that
the background is not transformed, $\dd\o_0=0$, as expected, and
$\dd\eta_1=D_0 \e_{\rm L}\ $,\  $\dd\r_1=\e_{\rm L}$ and, for 
$n\ge 2\ $,\break $\dd\eta_n=[\eta_{n-1},\e_{\rm L}]\ $,\
$\dd\r_n=[\r_{n-1},\e_{\rm L}]$, where $D_0$ is the covariant 
derivative with $\o_0$. Again we define 
$\eta'(\t)=\sum_{n=1}^\infty n\, \eta_n\, \t^{n-1}$ and
$\dh \eta(\t)=\sum_{n=1}^\infty (\d\eta_n)\, \t^n$, and idem 
for $\dh\r$. Then $\d\wt\s=\dh\eta-\eta'\D+\dh\r\D$ and for 
the curvature we find
\be\label{h6}
\Rt(\t)=\Rh(\t) + \xi(\t)\D
\ee
with
\ba\label{h7}
\Rh(\t) \equiv \Rh&=& R_0+\dh\eta+\o_0\eta+\eta\o_0+\eta^2
\nonumber
\\
\xi(\t)\equiv\xi&=& \dh\r+[\o_0+\eta\, ,\, \r]-\eta' \ .
\ea
Two useful identities which follow from the Bianchi 
identity $\d\Rt=[\Rt,\ot]$ are
\ba\label{h8}
\dh\Rh&=& [\Rh\, ,\, \o_0+\eta]
\nonumber
\\
\dh\xi&=& -\xi (\o_0+\eta)-(\o_0+\eta)\xi +[\Rh,\r]
-\dh\eta'-(\o_0+\eta)\eta'-\eta'(\o_0+\eta) \ .\ \
\ea

Finally, the cut-off gravitational 8-form $\wt X_8$ 
then is given by
\ba\label{h9}
\wt X_8 &=& {1\over 192 (2\pi)^3} 
\left( \tr\Rt^4-{1\over 4} (\tr\Rt^2)^2 \right)
\nonumber
\\
&=& {1\over 192 (2\pi)^3} 
\left( \tr\Rh^4 + 4\tr\Rh^3 \xi\ \D
-{1\over 4} (\tr\Rh^2)^2 - \tr\Rh^2\, \tr\Rh\xi\ \D \right) \ .
\ea
By construction, $\wt X_8$, as well as each of the four 
terms individually, is invariant under local Lorentz 
transformations. Hence, using $\Ct=C\t+B\D$ and 
$\D=\sum_\a \D_\a$, we find
\ba\label{h10}
\St_{\rm GS} &=& \St_{\rm GS}^{(1)} 
+ \sum_\a \St_{{\rm GS}}^{(2,\a)}
\nonumber
\\
\St_{\rm GS}^{(1)} &=& -{T_2\over 192 (2\pi)^4} 
\int C \left( \tr\Rh^4-{1\over 4} (\tr\Rh^2)^2 \right) \t
\nonumber
\\
\St_{{\rm GS}}^{(2,\a)} &=& -{T_2\over 192 (2\pi)^4} 
\int \Big( B\tr\Rh^4 + 4 C \tr\Rh^3\xi\, \t
-{1\over 4} B (\tr\Rh^2)^2 
- C \tr\Rh^2 \tr\Rh\xi\, \t \Big)\ \D_\a \ .
\nonumber
\\
\ea
While $\St_{\rm GS}^{(1)}$ involves an integral over all 
of $M_4\times X$, each $\St_{{\rm GS}}^{(2,\a)}$ reduces 
to an integral over $M_4\times Y_\a$. We could perfectly 
well do this reduction first and then compute the gauge 
variation of each $\St_{{\rm GS}}^{(2,\a)}$. However, as 
for the Yang-Mills case, the computations are 
more compact if we first take the variation and then 
evaluate the integral. Using 
$\dd C=\d\Lambda$, $\dd B=\Lambda$ and the invariance 
of the gravitational terms we find
\ba\label{h11}
\dd\St_{\rm GS}^{(1)} &=& -{T_2\over 192 (2\pi)^4} 
\int \d\Lambda \left( \tr\Rh^4-{1\over 4} (\tr\Rh^2)^2 \right) \t
\\
\label{h11b}
\dd\St_{{\rm GS}}^{(2,\a)} &=& -{T_2\over 192 (2\pi)^4} 
\int \Big( \Lambda\tr\Rh^4 + 4 \d\Lambda \tr\Rh^3\xi\, \t
-{1\over 4} \Lambda (\tr\Rh^2)^2 
- \d\Lambda \tr\Rh^2 \tr\Rh\xi\, \t \Big)\ \D_\a \ .
\nonumber
\\
\ea
We will discuss $\dd\St_{\rm GS}^{(1)}$ later on. As 
familiar by now, thanks to the presence of $\D_\a$,
each $\dd\St_{{\rm GS}}^{(2,\a)}$ is reduced to 
an integral over $M_4\times Y_\a$ with every $\t^k \D_\a$
contributing a factor ${1\over k+1}$. On $M_4\times Y_\a$, 
the derivative $\dh$ acts as an ordinary derivative $\d$,
and we are allowed to integrate by parts. As explained in 
eq. (\ref{h11a}) above, it is easy to see
that exactly the same result is obtained if
one first replaces $\d$ by $\dh$ and integrates all $\dh$ 
by parts 
directly in (\ref{h11b}), remembering that 
$\dh$ only acts on the $\o_0$, $\eta_n$ and $\r_n$, but 
not on the $\t^n$.
Using this observation, we rewrite
\be\label{h12}
\dd\St_{{\rm GS}}^{(2,\a)} = -{T_2\over 192 (2\pi)^4} 
\int  \Lambda \Big(\tr\Rh^4 - 4 \t\ \dh (\tr\Rh^3\xi)
-{1\over 4}  (\tr\Rh^2)^2 
+ \t\ \dh (\tr\Rh^2 \tr\Rh\xi) \Big)\ \D_\a \ .
\ee
Now use the identities (\ref{h8}) to show that $\dh \tr\Rh^2=0$
and
\be\label{h12a}
\dh\, (\tr\Rh^l\xi)=-\tr \Rh^l\left( 
\dh\eta'+(\o_0+\eta)\eta'+\eta'(\o_0+\eta) \right)
= -\, {1\over l+1}\, {\partial\over \partial\t} \tr\Rh^{l+1}
\ee
so that
\be\label{h13}
\dd\St_{{\rm GS}}^{(2,\a)} = -{T_2\over 192 (2\pi)^4} 
\int  \Lambda \left( 1 + \t   {\partial\over \partial\t} \right) 
\left( \tr\Rh^4 -{1\over 4}  (\tr\Rh^2)^2 \right)\ \D_\a \ .
\ee
Next, we expand the integrand in powers of $\t$. Writing
$\Rh=\sum_{n=0}^\infty \Rh_n\, \t^n$ with $\Rh_0=R_0$
and $\Rh_n=\dh\eta_n+\o_0\eta_n+\eta_n\o_0
+\sum_{r=1}^{n-1}\eta_r\eta_{n-m}$ for $n\ge 1$, 
we see that the 1 in the 
parenthesis in (\ref{h13}) contributes a factor
${1\over n+m+k+l+1}$ to the integral over $r_\a$
while the $\t {\partial\over \partial\t}$
contributes a factor ${n+m+k+l\over n+m+k+l+1}$, 
both adding up to 1. As a result, we get
\be\label{h14}
\dd\St_{{\rm GS}}^{(2,\a)} = -{T_2\over 192 (2\pi)^4} 
\int_{M_4\times Y_\a}  \Lambda 
\left( \tr R^4 -{1\over 4}  (\tr R^2)^2 \right) 
\equiv -{T_2\over 2\pi} \int_{M_4\times Y_\a} \Lambda\  X_8(R)\ ,
\ee
where now
\be\label{h15}
R=\sum_{n=0}^\infty \Rh_n \ .
\ee
Clearly, $R$ is the value of the curvature ``in the bulk''
of $M_4\times X$  (or its 
appropriate pullback onto $M_4\times Y_\a$), 
and corresponds to the fluctuating 
geometry $\o=\o_0+\s$ without cutting off anything. 
Hence we see that, in the end, this rather sophisticated 
treatment reproduces the same result as the more naive 
$\St_{\rm GS}'=-{T_2\over 2\pi}\int \Ct\w X_8$ of eq (\ref{g2}). 
It is clear from our analysis that this same simplification
occurs for any 
invariant quantity made from combinations of $\tr R^l$\,
or $\tr F^k$.

It remains to discuss $\dd\St_{\rm GS}^{(1)}$. With 
$\Lambda=\e_i\, \o_i$, the integral will be non-vanishing 
only if $\tr \Rh^4 -{1\over 4}  (\tr \Rh^2)^2$ is a 
3-form on $M_4$ and a 5-form on $X$. If the fluctuations 
of the metric preserve the product structure $M_4\times X$, 
this is clearly impossible, and we conclude 
\be\label{h16}
\dd\St_{\rm GS}^{(1)}=0\ .
\ee 
For more general fluctuations, 
however, it is less clear what happens and we will not pursue 
this issue further.

\section{Conclusions}
\setcounter{equation}{0}

We have reconsidered the anomaly cancellation mechanism
on $G_2$-holonomy manifolds with conical singularities, 
first outlined 
in \cite{Witten}. It turned out that we needed to 
modify the eleven-dimensional Chern-Simons and 
Green-Schwarz terms, and similarly the interactions 
$S_1$ and $S_2$ present on ADE singularities, by (smoothly)
cutting 
off the fields close to the conical singularities. This 
induces anomalous variations of the  cut-off 3-form 
field $\Ct$ and of the cut-off non-abelian gauge field 
$\At$. These anomalous variations are localized in the 
regions close to the conical singularities where the 
cut-off is done. This implies that the corresponding 
non-invariance of the action is also localized there 
and we get one $\dd S^{(\a)}$ term for each conical 
singularity $P_\a$. 
Each of these terms then exactly cancels the various
anomalies that are present at these singularities 
due to the charged chiral fermions living there.
Thus anomaly cancellation indeed occurs locally, i.e. 
separately for each conical singularity.

Throughout the whole discussion it is always assumed that the 
$G_2$-holonomy mani\-fold is compact, although the explicit
examples of conical singularities are actually taken from 
the known non-compact $G_2$-holonomy manifolds, assuming that 
conical singularities on compact $G_2$-manifolds have the
same structure. As mentioned in the introduction, there 
exist close relatives of $G_2$-holonomy manifolds which 
are {\it weak} $G_2$-holonomy manifolds. In this case, 
it is quite easy to construct {\it compact} examples with 
conical singularities and explicitly known metrics. This 
is done in \cite{BM} and will be briefly 
recalled in the appendix. The conical singularities are 
exactly as assumed in the present paper, namely for 
$r_\a\to 0$ they are cones on some $Y_\a$ with the same 
$Y_\a$'s as considered here. This implies that the whole 
discussion of chiral fermions present at the singularities 
and of the anomaly cancellation of the present paper 
directly carries over
to these weak $G_2$-holonomy manifolds.

\vskip 10.mm
\noindent
{\bf\large Acknowledgements}
\vskip 3.mm

\noindent
Steffen Metzger gratefully acknowledges support
by the Gottlieb Daimler- und Karl Benz-Stiftung.
We would like to thank Luis Alvarez-Gaum\'e, 
Jean-Pierre Derendinger, Jean Iliopoulos,
Ruben Minasian, Ivo Sachs, Julius Wess
and Jean Zinn-Justin for helpful discussions.
 \vskip 2.cm

\section{Appendix}
\renewcommand{\theequation}{A.\arabic{equation}}
\setcounter{equation}{0}

Here we will briefly recall the geometry of the singular 
weak $G_2$-holonomy manifolds $X$ constructed in 
\cite{BM}. Although they have weak $G_2$-holonomy 
rather than $G_2$-holonomy, they are the prototype 
of the compact manifolds with conical singularities 
one has in mind throughout the present paper. 

In \cite{BM} it was shown that for 
every non-compact $G_2$-manifold that is asymptotic 
(for large $r$) to a cone on $Y$, there is an associated 
compact weak $G_2$-manifold with its metric given by
\be\label{a1}
\d s^2_{X} = \d r^2 + \left( R \sin{r\over R}\right)^2 
\, \d s^2_Y \ , \quad 0\le r\le \pi R \ .
\ee
It has two conical singularities. The first one, at 
$r=0$, is a cone on $Y$, while the second one, at 
$r=\pi R$, is a cone on $-Y$. Here $-Y$ equals $Y$ 
but with its orientation reversed. 
This reversal of orientation simply occurs 
since we define $Y_\a$ always  such that the normal 
vector points away from the singularity.
Hence:
\be\label{a2}
Y_1=Y \ , \quad Y_2=-Y \ .
\ee

For these examples we have all the necessary global 
information, and it was shown in \cite{BM} that the 
square-integrable harmonic $p$-forms on $X$, for 
$p\le 3$, are the trivial extensions of the 
square-integrable harmonic $p$-forms on $Y$. In particular,
we have $b^1(X)=b^1(Y)=0\, $, $\ b^2(X)=b^2(Y)$ and 
$b^3(X)=b^3(Y)$.

According to the general cut-off procedure described 
in section 3.1, for these examples one introduces 
two local coordinates $r_1=r$ and $r_2=\pi R-r$. 
It follows that $\D=\D_1+\D_2$, where, in the limit 
of vanishing regularisation,
\ba\label{a3}
\D_1&=&\dd(r_1-\rt)\, \d r_1= \dd(r-\rt)\, \d r 
\nonumber
\\
\D_2&=&\dd(r_2-\rt)\, \d r_2= -\, \dd(r-(\pi R-\rt))\, \d r \ .
\ea
Then for a smooth 10-form $\f$ one has
\be\label{a4}
\int_{M_4\times X}\ \f\ \D 
=\int_{M_4\times Y}\ \f\Big\vert_{r=\rt} \
-\int_{M_4\times Y}\ \f\Big\vert_{r=\pi R-\rt} \
=\int_{M_4\times Y_1}\ \f +\int_{M_4\times Y_2}\ \f \ .
\ee

\vskip 2.cm



\end{document}